# A Practical Approach to Managing Spreadsheet Risk in a Global Business

Tom Lemon, Ewen Ferguson
Protiviti Limited, Grand Buildings, 1-3 Strand, London WC2N 5AB
t.lemon@protiviti.co.uk

## ABSTRACT

*Spreadsheets are used extensively within today's organisations. Although spreadsheets have many benefits, they can also present a significant risk exposure, requiring appropriate management. Protiviti has worked with a number of organisations, ranging in size up to huge multi-nationals, to help them build appropriate spreadsheet governance frameworks, including the design and implementation of policies, minimum design standards, control processes, training and awareness programmes and the consideration and implementation of spreadsheet management tools. This paper presents a case-study explaining the practical and pragmatic approach that was recently taken to control spreadsheet risk at one of Protiviti's clients – a global energy firm (referred to as 'client' below).*

## 1   INTRODUCTION

We all know that today's businesses place a lot of reliance on spreadsheets. They have been doing so for many years now. In that time, spreadsheet technology has evolved, giving us more features that provide even more flexibility and functionality. As a result, spreadsheets continue to be incredibly useful tools that have become an irreplaceable component of many business processes. They will never go away. It would be wrong to suggest that they should (at least in their entirety).

We've also seen the introduction of new regulations over the last decade that have in many cases helped raise the awareness of spreadsheet risk to executive management (e.g. the Sarbanes-Oxley Act) [McCready and McGouran, 2009]. Additionally, organisations have come under increased scrutiny from industry regulators to prove that they have adequate controls around critical End User Computing applications and that they themselves are free from errors [FINRA Notice of Examination, 2010]. Another example is that the UK Financial Services Authority has recently been fining organisations, with reasons cited including reliance on spreadsheets [FSA Final Notice to Credit Suisse, 2008].

Companies have consequently invested a great deal of time and money to improve their internal control frameworks, although few really fully understand what the risk exposure is and what they need to do about it. So, where has this left spreadsheets? What, in practice, has been done to help mitigate the risk of spreadsheet errors in organisations?

Protiviti has worked with many organisations to tackle this problem. In our experience, we have seen that companies are increasingly aware of the issue through a combination of actual errors occurring and the encouragement of auditors and regulators to address accounting and legislative requirements. This has without doubt helped. We have also seen that given the recent economic climate, organisations are trying to cut costs and achieve an acceptable level of control that is pragmatic and cost effective. Finding this balance is often a challenge and there is certainly not a one-size-fits-all approach for all organisations. The right solution will depend on many factors, including (in no particular order):

- Prevalence of spreadsheets in use;

- The type of business processes that rely on spreadsheets (e.g. finance, operations, management reporting, core business);

- Existence of expert spreadsheet developers and/or support teams (e.g. Rapid Application Development (RAD) teams);

- Level of spreadsheet expertise in the user community;

- Materiality or significance of spreadsheets being used;

- Company culture and risk appetite;

- Existing effectiveness and awareness of internal control; and

- Many, many more…

This paper provides a case study summarising how Protiviti helped one client approach spreadsheet risk management recently. The client is a large, global organisation with energy production, trading, marketing, middle office and finance operations and relies on thousands of spreadsheets to perform significant tasks across a variety of business critical processes.

## 2   THE ISSUE

Spreadsheets and other End User Computing (EUC) applications (e.g. Access databases) were being used by our client to perform many important financial and operational processes. Hundreds of applications were identified that had a significant direct or indirect impact on financial reporting and hundreds more were identified that were being relied upon to support key operational processes. It was recognised that errors in these applications could result in considerable loss to the organisation through inappropriate decision making or missed opportunities, as well as presenting a regulatory and reputational risk should an error result in a material financial reporting misstatement. The issue was accentuated by the fact that thousands of less significant spreadsheets were also known to be used that could in aggregate cause similarly significant problems to our client.

Audits have shown that nearly all spreadsheets contain errors [Panko and Ordway, 2005]. As is so often the case with this area of risk, it usually requires something to go wrong for the right people to take notice and for something to be done. This was no exception. A number of spreadsheet errors occurred over a few years that gradually caught the attention of senior executives and a group wide mandate was given to get it fixed.

## 3 THE SOLUTION

### 3.1 Overview

Our client had two areas of focus:

1. Financial EUC applications that had a significant direct or indirect impact on financial reporting, across all parts of the business; and

2. Operational EUC applications that had a significant potential for operational loss through inappropriate decision making or missed opportunities, in the trading area of the business.

As you would expect, there were different requirements and different stakeholders in each of these areas of focus; however, largely the same approach was taken to address the issue across both.

Our client established a spreadsheet control framework. This is the structure that an organisation implements to define the spreadsheet risks and the associated controls that should be considered [Protiviti, 2008]. Our client's control framework was structured as follows:

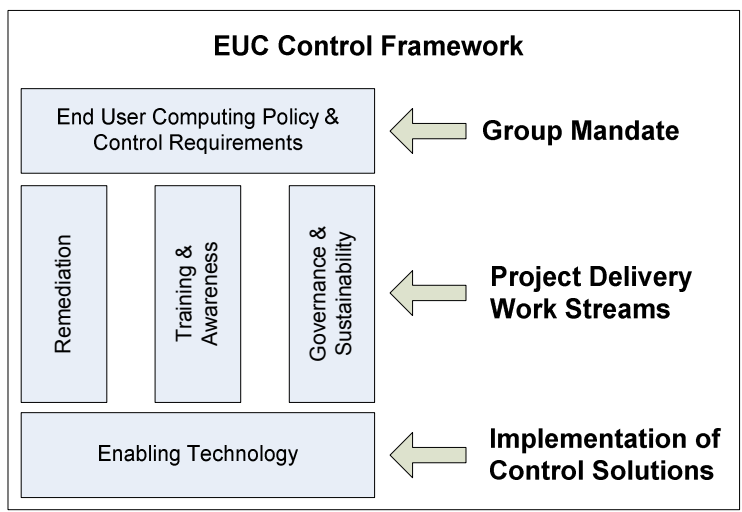

The solution included a Group mandate (described below) and a project to deliver three streams of work: remediation, training and awareness and governance and sustainability. Enabling technologies were also implemented as part of the project to support some aspects of the control framework.

The following sections describe each component of the solution in further detail.

### 3.2 End User Computing Policy & Control Requirements

Our client's Group Control function established an EUC framework, which consisted of a policy setting out key definitions and control requirements for specific categories of spreadsheet. The control requirements were broken down into the following areas:

**Inventory Management**

An exercise was undertaken to identify all Financial EUC applications – largely consisting of Excel spreadsheets – across the entire Group and all Operational EUC applications in the trading area of the business.

The EUC Policy required that these inventories be kept up-to-date and maintained. However, spreadsheets are created daily. They also get replaced or retired almost as frequently. It's a constantly moving target. Inventory maintenance has therefore long been problematic for large organisations. A common issue is that the responsibility for inventory management is segregated from the community of people that actually use them. It's important to try and bridge this gap as effectively as possible, whilst maintaining control of the corresponding data. This can be achieved by improving awareness through better communications or by empowering line managers and / or owners of applications to be accountable for this activity in their respective business areas.

A successful solution at this client was the implementation of an online inventory application that enabled the owner community to register new EUC applications quickly and easily and update the details for existing ones, should they change. It also enabled line managers to quickly see what applications were recognised as being used to support the business processes for which they were accountable. This, combined with targeted training and awareness campaigns, and ongoing monitoring by internal control and operational risk functions, improved the effectiveness of their inventory management considerably.

**Design standards**

A selection of design standards were established and mandated for the most significant categories of EUC application. Standards were written for Excel spreadsheets and Access databases, although, as we've seen in most organisations, Excel spreadsheets were the more prevalent. The design standards covered the following principles:

- Improved documentation [Payette, 2006] – achieved by requiring the completion of standard documentation templates in all applications and including appropriate commentary to explain complex calculations and VBA code;

- Transparency of information – achieved by making data and calculations visible and clearly understood;

- Clear labelling – achieved by ensuring key data inputs, calculations, outputs, assumptions and units of measure are all adequately labelled;

- Separation of inputs, calculations and outputs – achieved through a combination of structural separation and visual formatting and labelling; and

- Critical cell locking – achieved by locking all cells with critical formulas and static data and activating worksheet protection.

**Validation processes**

All EUC applications were required to be independently validated prior to their first use and periodically thereafter. The validation process included checks for design logical integrity issues (e.g. formula inconsistencies, potential errors).

**Input, calculation and output checking requirements**

All EUC applications were required to include processes and techniques to check:

- The completeness and accuracy of inputs – a combination of check cells such as control totals, or criteria tests, and manual review techniques were built into all spreadsheets where appropriate; and

- The accuracy of calculations and outputs – a combination of automated checks and balances, manual reasonableness checks and variance analyses were also built into all spreadsheets where appropriate.

A standard review log template was also included in all spreadsheets to capture the results of manual checks and to evidence the performance of these checks on a periodic basis.

**Change management**

Changes to EUC applications were required to be properly documented, including the reason and rationale behind each change. Furthermore, every functional or structural change was required to be independently reviewed to ensure all control standards continued to be met and that no errors were introduced to functionality. This was evidenced initially through the inclusion of a standard change log template within every file.

Even so, history tells us that a manual change management approach is highly susceptible to breaking down over time. Reliance is placed entirely on spreadsheet users and owners to follow the process and there is no simple way to ensure that this is done. With this in mind, our client made the decision to evaluate and select a change management solution to provide additional rigour around this control for the most critical spreadsheets. ClusterSeven was selected and implemented for this purpose. The 'Enabling Technologies' section below outlines how this was deployed.

**Security**

EUC applications were required to be protected by a file password in appropriately restricted network directories. Our client recognised that Excel passwords are reasonably straightforward to crack and that they didn't therefore provide an optimum level of control. However, their main objective was to signify the intent that files should only be accessed by authorised personnel, which this achieved.

**Archiving**

Finally, EUC applications were required to be periodically archived and named using appropriate conventions to support version control.

### 3.3 Remediation

One of the project delivery work streams was: 'Remediation'. The objective of this work was to review all registered EUC applications that were deemed the most critical. This included approximately 700 Financial EUC applications and approximately 200 Operational EUC applications. Our remediation approach consisted of the following steps:

1. Diagnosis – an initial review of the EUC application against the control requirements defined previously. This gap analysis generated a file specific remediation plan.

2. Remediation – all control requirements were implemented to address identified gaps. The remediation extent and approach varied depending on the design of the existing file, user preference and the file's purpose (amongst other things). However, the control principles were consistently applied.

3. QA – an independent quality assurance activity was performed on all remediated files to validate that all control requirements had been met and to test that no functionality had been broken during the process. In addition, a series of design integrity tests [Pryor, 2004] were performed using a spreadsheet analysis tool to look for and address any inconsistencies and potential errors that existed in the file.

4. Transition – a transition meeting (or series of meetings) was held with the owner of the remediated file and any other key users to ensure they were comfortable with the implemented controls and understood their ongoing obligations to maintain them.

The output of this remediation cycle was a file that complied with all defined EUC control requirements. On average, this took between 3-5 man days of effort per file, depending on the application's complexity and its original state, however it should be noted that in our experience every organisation's requirements are different and therefore the time varies significantly from one company to the next.

### 3.4 Training & Awareness

Our client recognised that training and awareness was key to the ongoing sustainability of the EUC control framework [Protiviti, 2008]. As such, a number of different training programmes were piloted and implemented across the user population. These included:

- Computer Based Training (CBT) – an intensive CBT course was implemented that was designed for the owners and the developers of all critical spreadsheets. The course used the latest technology to step through all control requirements and provide extensive examples illustrating how they could be achieved. Our client considered the CBT a highly successful and helpful campaign that significantly improved the awareness and capability of a large number of people.

- Classroom Training – a classroom training course was piloted that provided a more hands-on version of the content included in the CBT course. This was designed for those people that preferred this method of learning and was therefore rolled out on a smaller scale to supplement the CBT.

### 3.5 Governance & Sustainability

The objective of this project work stream was to ensure the EUC control framework could be sustained over time and to put in place governance practices to provide oversight. This encompassed a number of activities that have been discussed elsewhere in this paper, in particular the consideration of the most appropriate inventory management process and enabling technologies, such as spreadsheet change management solutions.

### 3.6 Enabling Technology

The ClusterSeven Enterprise Spreadsheet Management (ESM) tool [Baxter, 2008] was selected by our client to provide more rigorous change control for the most critical spreadsheets. This was deployed in a subsequent phase of the project. The deployment approach consisted of the following key activities:

1. Reviewing the spreadsheet – to ensure all control requirements continued to be met (updating where required). It is important to establish a baseline before controlling spreadsheets. There is little point in controlling a spreadsheet that is not working in the first place [Ferguson, 2009].

2. Configuring ClusterSeven – to activate the monitoring of the file in the system and to define what types of changes should trigger alerts, requiring an independent review and sign off.

3. Reviewer training – to provide identified 'Reviewers' with the capability to use ClusterSeven to review and sign off exceptional changes.

## 4 CONCLUSION

Spreadsheets are here to stay and are in many cases critical business applications. Critical business applications need to have adequate controls. As well as meeting regulatory requirements, spreadsheet control helps to reduce potential losses due to errors and can introduce productivity and efficiency gains.

This case study presents a pragmatic approach to managing spreadsheet risk within a large global company. The approach provides a solid basis that can be scaled and customised to meet the requirements of different organisations. It is, however, vital to define objectives when embarking on a spreadsheet risk management initiative.